# Ferromagnetic properties of SrRuO$_3$ thin films deposited on the spin-triplet superconductor Sr$_2$RuO$_4$

Yusuke Sugimoto[1], Muhammad Shahbaz Anwar[1, *], Seung Ran Lee[2,3],
Yeong Jae Shin[2,3], Shingo Yonezawa[1], Tae Won Noh[2,3], and Yoshiteru Maeno[1]

[1]*Department of Physics, Graduate School of Science, Kyoto University, Kyoto 606-8502, Japan*
[2]*Center for Correlated Electron Systems, Institute for Basic Science (IBS), Seoul 151-747, Republic of Korea*
[3]*Department of Physics and Astronomy, Seoul National University, Seoul 151-747, Republic of Korea*

**Abstract**

We report magnetic properties of epitaxial thin films of the itinerant ferromagnet SrRuO$_3$ deposited on the cleaved *ab* surface of the spin-triplet superconductor Sr$_2$RuO$_4$. The films exhibit ferromagnetic transition near 160 K as in the bulk SrRuO$_3$, although the films are under 1.7% compressive strain. The observed magnetization is even higher than that of the bulk SrRuO$_3$. In addition, we newly found that the magnetization relaxation after field removal is strongly anisotropic: two relaxation processes are involved when magnetic domains are aligned along the *ab*-plane.

## 1 Introduction

The proximity effect between a ferromagnet (FM) and a spin-triplet superconductor (TSC) has been theoretically explored expecting novel superconducting phenomena [1-3]. In particular, the penetration of the superconducting order parameter of a chiral *p*-wave TSC into a FM is predicted to be largely dependent on the relative angle between the *d*-vector of the TSC and the magnetization of the FM [1]. Experimental investigations, however, are still lacking.

SrRuO$_3$ (SRO113) is an itinerant FM metal, whose Curie temperature ($T_{\text{Curie}}$) in bulk is about 160 K [4]. Among the Ruddlesden-Popper series Sr$_{n+1}$Ru$_n$O$_{3n+1}$, SRO113 is an infinite-layer material ($n = \infty$) while Sr$_2$RuO$_4$ (SRO214) is the $n = 1$ member. In SRO113, although the RuO$_6$ octahedral rotation leads to orthorhombic crystal structure, the pseudocubic lattice constant $a = 3.930$ Å is still useful [5]. Considering relevant atomic orbitals, the energy levels of the five-fold degenerate Ru 4*d* orbitals split into two $e_g$ and three $t_{2g}$ levels. A large crystal field yields the low-spin configuration of $S = 1$ for Ru$^{4+}$; its four electrons occupy only the $t_{2g}$ levels. A density-functional calculation estimates the magnetic moment (*M*) of SRO113 in the ground state is below 2 $\mu_B$ per Ru ion [7] for the actual orthorhombic structure, close to experimental values of saturated moments ($\mu_{sat}$ = 1.0-1.8 $\mu_B$/Ru$^{4+}$) [6]. For the hypothetical cubic structure, it is 1.17 $\mu_B$/Ru$^{4+}$ [7].

---

[*] *Corresponding author M. S. Anwar; anwar@scphys.kyoto-u.ac.jp*

*M* as well as $T_{Curie}$ of SRO113 thin films have been known to be affected by subtle changes of strain which can be varied by choosing different single-crystalline substrates [5]. For example, SRO113 deposited on SrTiO$_3$ under tensile strain of 0.4% shows $T_{Curie} \approx$ 150 K and $\mu_{sat} \approx$ 1.4 $\mu_B$/Ru$^{4+}$ [5].

SRO214 is most probably a chiral *p*-wave TSC with the transition temperature $T_c \approx$ 1.5 K [8,9]. Recently, SRO214 has been attracting further attention for the non-trivial topological nature of its superconducting wave function [8], which can be investigated with junction techniques [8-12]. Therefore, device fabrication using SRO214 is of great interest. Despite such demands, it is challenging to obtain a superconducting SRO214 film exhibiting superconductivity except for one report [13] with $T_c \approx$ 1 K. As an alternative approach to a FM/TSC junction, we succeeded in fabricating SRO113/SRO214 hybrids with electrically conductive interface [14] using SRO214 single crystal as a substrate instead of using SRO214 films. In this study, we present the magnetic properties of the SRO113/SRO214 hybrids. Our studies on this hybridized system with TSC may help opening up a new research field of unconventional proximity effects.

## 2 Experimental

We grew high quality SRO214 single crystals using a floating zone method [15]. Depending on growth parameters, SRO214 single crystals may contain Ru-metal, Sr$_3$Ru$_2$O$_7$ or SRO113 inclusions. Thus, before growing films, we very carefully examined the SRO214 substrates by X-ray diffraction. In addition, under optical microscope the Ru-metal and Sr$_3$Ru$_2$O$_7$ inclusions are examined, and then using a superconducting quantum interference device (SQUID) magnetometer (Quantum Design MPMS-XL), the SRO113 impurities in the substrate are quantified.

To prepare SRO214 substrates, we cut SRO214 crystals along the *ac*-plane and cleaved them along the *ab*-surface with the approximate size of 3×3×0.5 mm$^3$. SRO113 thin films are grown on the cleaved *ab*-surface using a pulsed laser deposition (PLD) technique. Details of the film growth are described in Ref [14]. The crystallographic quality of the films is inspected with an X-ray diffractometer with the Cu Kα radiation ($\lambda$ = 1.5406 Å).

Magnetic properties of the hybrids have been investigated using the SQUID magnetometer. The temperature dependence of the magnetization is measured under various fields after field cooling with the fields parallel to either the *a*-axis (in-plane) or *c*-axis (out-of-plane) from room temperature to 4 K. The magnetization loop between −6 T and 6 T is obtained at 4 K after zero-field cooling. The magnetization of the film is obtained by subtracting the substrate contribution. We also measure the time dependent relaxation of remnant magnetization at zero field and 4 K after cooling under 3 T and then switching off this field. For this relaxation study, we acquired data at every 5 minutes without changing temperature and field conditions. To minimize the residual field, we utilized the superconducting signal (*in-situ*) of lead to detect the exact value of the field, by placing a lead spherical sample on the sample rod 5-cm away from the SRO113/SRO214 sample.

## 3 Results

Figure 1 represents a X-ray diffraction pattern of a 15-nm-thick SRO113 film on SRO214 in the range 40° < 2θ < 50°. In this range scattering-angle there are two independent peaks, which are the

(006) peak of SRO214 and the (002) peak of SRO113. The peak position of the SRO113 film ($2\theta$ = 45.32°) appears at lower angle compared to the bulk SRO113 ($2\theta$ = 46.15° from $c$ = 3.932 Å, the blue solid line) [5]. It indicates that the $c$-axis of the SRO113 film is elongated by about 1.7%. For comparison, the lattice mismatch at 300 K between the bulk SRO113 ($a_{113}$ = 3.930 Å) and SRO214 ($a_{214}$ = 3.871 Å) is −1.5%. Thus, the $c$-axis elongation indicates that our SRO113 film is compressively strained. Thickness fringes around the SRO113 (002) peak indicate that our film and SRO113/SRO214 interface are atomically flat.

Figure 2(a) presents magnetization loops of a 50-nm-thick SRO113/SRO214 film for fields along the $c$-axis and the $ab$-plane of SRO214 including the contribution from this substrate. Although the paramagnetic contribution of the substrate dominates the data, one can clearly see ferromagnetic component at lower fields. To remove the substrate contributions, we deduced the slope of the linear paramagnetic part at higher fields between 4 T to 6 T. Then we subtracted the linear contribution. After subtracting this contribution, we obtain the ferromagnetic loop of the film plotted in Fig. 2(b). Note that saturation magnetization along the $c$-axis is higher than that along the $a$-axis. This magnetic anisotropy is not much different from SRO113 thin films deposited on other substrates. Figure 2(c) shows the temperature dependence of the magnetization in various fields. Figure 2(d) represents the remanent magnetization of the film at $H = 0$ obtained by subtracting the substrate contribution. Interestingly, $T_{\text{Curie}}$ of the film is about 160 K, almost the same as that of bulk in spite of the relatively strong strain of the film. Surprisingly, the remnant magnetization for $T \to 0$ is about 3.5 $\mu_\text{B}$/Ru$^{4+}$, which is substantially larger than 2 $\mu_\text{B}$/Ru$^{4+}$, the upper limit of the low-spin state of the Ru$^{4+}$ ion under octahedral coordination with assumption that the orbital moment is quenched.

To further characterize the magnetization, we investigated the relaxation of magnetization $M(t)$. We cooled down a 30-nm thick SRO113/SRO214 thin film under 3 T (higher than the coercive field) and switched off the field (the residual field of MPMS was also removed) at 4 K. Figure 3 shows the normalized relaxation curves $m(t) = M(t)/M(0)$, revealing a remarkable anisotropy in the relaxation process. For $H \parallel c$, the relaxation is rather slow and weak, and well fitted by the single exponential function $m(t) = m_0 \exp(-t/\tau_0) + (1 - m_0)$ with $m_0 = 0.03 \pm 0.001$ and $\tau_0 = 1290 \pm 50$ min. This relaxation process is perhaps mainly due to domain-wall motion accompanied by magnetization flipping from the $M//c$ to $M//\text{-}c$ direction. In contrast, the double exponential function $m(t) = m_1 \exp(-t/\tau_1) + m_2 \exp(-t/\tau_2) + (1 - m_1 - m_2)$ is necessary to fit the $m(t)$ data with $H // ab$. We obtain the parameters $\tau_1 = 510 \pm 40$ min., $\tau_2 = 98.4 \pm 4$ min., $m_1 = 0.26 \pm 0.006$, and $m_2 = 0.22 \pm 0.01$. The longer relaxation process with characteristic time $\tau_1$ may correspond to the decrease in $M//a$ component and increase in $M//\text{-}a$; the shorter relaxation process with $\tau_2$ to the increase in the easy-axis direction $M//c$.

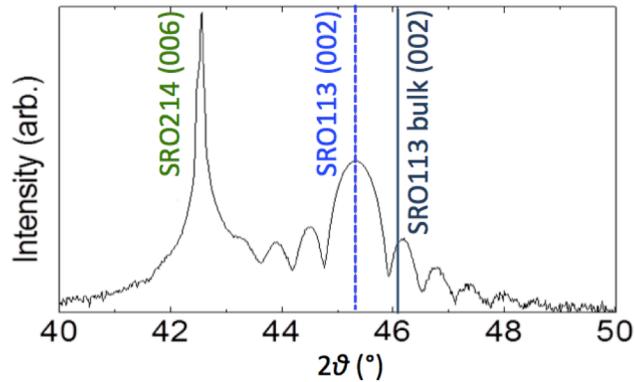

**Figure 1**: X-ray diffraction pattern of a 15-nm-thick SRO113/SRO214. The vertical axis is in the logarithmic scale. The dotted line represents the position of the (002) peak of the film and solid line shows the corresponding position of the bulk SRO113.

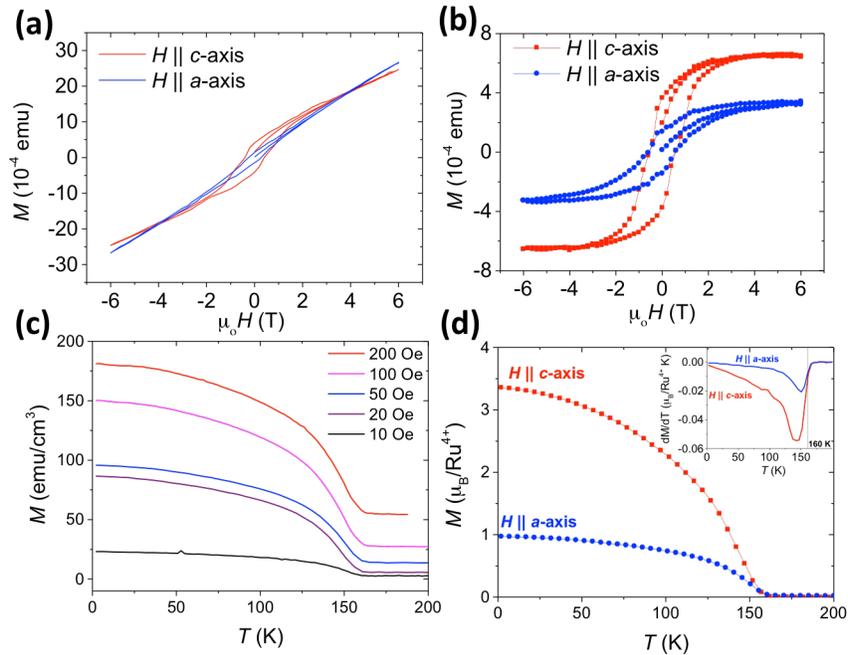

**Figure 2**: (a) Magnetization loops of a 50-nm-thick film of SRO113/SRO214 with fields along the *a*-axis and the *c*-axis before subtracting the substrate contribution. (b) Magnetization loop after subtracting the linear paramagnetic signal from the substrate. (c) Temperature dependence of the magnetization of the hybrid with a 50 nm film with various fields along the *a*-axis. (d) Temperature dependence of the magnetization of the film with zero field after 3 T cooling along the *a*- and *c*-axis. The substrate contributions have been subtracted. The inset shows temperature derivative of the magnetization as a function of temperature, which shows the Curie temperature is about 160 K (solid vertical line).

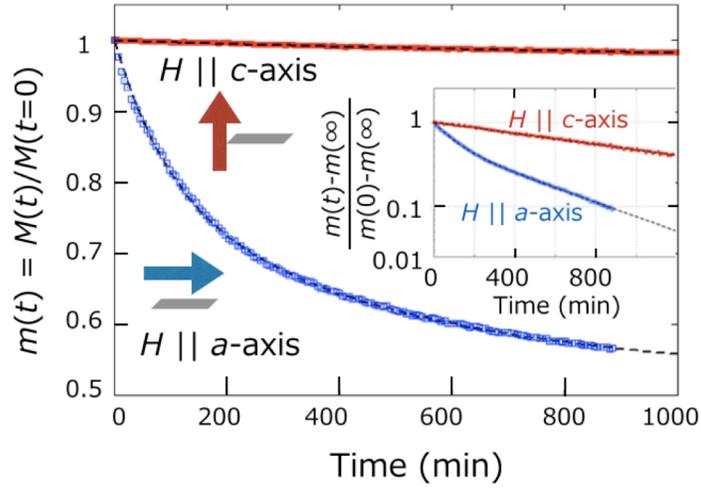

**Figure 3**: Normalized magnetization as a function of time for a 30-nm thick SRO113/SRO214 film. The measurements are performed at zero field and 4 K after cooling from room temperature under 3 T. The dotted curves are results of fittings of exponential functions to the data (see text). The inset represents $\{m(t)-m(\infty)\}/\{m(0)-m(\infty)\}$ in semi-log scale.

## 4 Summary

We investigated ferromagnetic properties of SRO113 thin films deposited on SRO214 spin-triplet superconductor substrates. The films are under sever compressive strain that results in an elongation of the $c$-axis by up to 1.7%. We confirmed our earlier report [14] that SRO113/SRO214 films have extremely high saturated magnetization (3.5 $\mu_B$/Ru$^{4+}$) along the $c$-axis. The magnetization loops also confirm strong anisotropy with the easy-axis along the out-of-plane $c$-direction. Correspondingly, the magnetization relaxation for the initial magnetization along the $c$-axis is weak and slow, whereas along the $a$-axis it is strong and faster with clear two relaxation components. We infer that such double-exponential relaxation is due to the re-arranging processes of the ferromagnetic domains. Our work will serve as a basis for the study of spin-triplet superconducting proximity effect.

## Acknowledgement

We acknowledge technical support from C. Sow and A. Ohtsuka. This work was supported by the "Topological Quantum Phenomena" (No. 22103002) KAKENHI on Innovative Areas from the Ministry of Education, Culture, Sports, Science and Technology of Japan (MEXT) and is also supported by the Institute for Basic Science (IBS) in Korea. MSA is supported as an International Research Fellow of the Japan Society for the Promotion of Science (26-04329).